\documentstyle[epsf]{aipprocsp}

\def\what{\widehat}

\def\To{\Rightarrow}

\def\beq{\begin{equation}}
\def\eeq{\end{equation}}

\def\chitil{\wtil \chi}
\def\mchitil{m_{\chitil}}

\def\eg{{\it e.g.}}
\def\etal{{\it et al.}}

\def\sq{\wt q}

\def\hsm{h_{\rm SM}}
\def\mhsm{m_{\hsm}}
\def\hl{h^0}
\def\hh{H^0}
\def\ha{A^0}
\def\hp{H^+}
\def\hm{H^-}
\def\hpm{H^{\pm}}

\def\mhh{m_{\hh}}
\def\mha{m_{\ha}}

\def\mhpm{m_{\hpm}}
\def\tanb{\tan\beta}

\def\mt{m_t}

\def\mz{m_Z}
\def\mw{m_W}

\def\chitil{\wt\chi}
\def\cnone{\wt\chi^0_1}

\def\snu{\wt\nu}

\def\msnu{m_{\snu}}
\def\mcnone{m_{\cnone}}

\def\h{h}
\def\mh{m_{\h}}
\def\wt{\widetilde}

\def\cpmone{\wt \chi^{\pm}_1}

\def\mcpmone{m_{\cpmone}}

\def\dmm{\Delta^{--}}
\def\mdmm{m_{\dmm}}

\def\MPL #1 #2 #3 {Mod.~Phys.~Lett.~{\bf#1},\  #2 (#3)}
\def\NPB #1 #2 #3 {Nucl.~Phys.~{\bf#1},\  #2 (#3)}
\def\PLB #1 #2 #3 {Phys.~Lett.~{\bf#1},\  #2 (#3)}
\def\PR #1 #2 #3 {Phys.~Rep.~{\bf#1},\ #2 (#3)}
\def\PRD #1 #2 #3 {Phys.~Rev.~{\bf#1},\  #2 (#3)}
\def\PRL #1 #2 #3 {Phys.~Rev.~Lett.~{\bf#1},\  #2 (#3)}
\def\RMP #1 #2 #3 {Rev.~Mod.~Phys.~{\bf#1},\  #2 (#3)}
\def\ZP #1 #2 #3 {Z.~Phys.~{\bf#1},\  #2 (#3)}
\def\IJMP #1 #2 #3 {Int.~J.~Mod.~Phys.~{\bf#1},\  #2 (#3)}
\def\mVV{M_{VV}}

\def\hpm{H^{\pm}}

\def\call{{\cal L}}

\def\lam{\lambda}
\def\br{BF}

\def\gam{\gamma}
\def\sigrts{\sigma_{\tiny\rts}^{}}

\def\etal{{\it et al.}}

\def\sighbar{\overline \sigma_{\h}}

\def\anti{\overline}
\def\epem{e^+e^-}
\def\zstar{Z^\star}
\def\wstar{W^\star}

\def\mupmum{\mu^+\mu^-}

\def\rts{\sqrt s}
\def\ie{{\it i.e.}}
\def\eg{{\it e.g.}}

\def\anti{\overline}

\def\mw{m_W}
\def\mz{m_Z}
\def\h{h}
\def\mh{m_{\h}}
\def\gamh{\Gamma_{\h}^{\rm tot}}
\def\hsm{h_{SM}}
\def\mhsm{m_{\hsm}}
\def\gamhsm{\Gamma_{\hsm}^{\rm tot}}
\def\tanb{\tan\beta}
\def\hl{h^0}

\def\ha{A^0}
\def\mha{m_{\ha}}

\def\hh{H^0}
\def\mhh{m_{\hh}}

\def\fbi{~{\rm fb}^{-1}}

\def\mev{~{\rm MeV}}
\def\gev{~{\rm GeV}}
\def\tev{~{\rm TeV}}

\def\mt{m_t}

\def\overlay#1#2{\ifmmode \setbox 0=\hbox {$#1$}\setbox 1=\hbox to\wd 0{\hss
$#2$\hss }\else \setbox 0=\hbox {#1}\setbox 1=\hbox to\wd 0{\hss #2\hss }\fi
#1\hskip -\wd 0\box 1}
\def\case#1/#2{{\textstyle{#1\over#2}}}
\def\9{\phantom 0}      %%% for lining up numbers in columns
\renewcommand\linebreak{\unskip\break} %% breaks line & still justifies
\newcommand{\alt}{\mathrel{\raisebox{-.6ex}{$\stackrel{\textstyle<}{\sim}$}}}
\newcommand{\agt}{\mathrel{\raisebox{-.6ex}{$\stackrel{\textstyle>}{\sim}$}}}
\def\lsim{\alt}
\def\gsim{\agt}
\catcode`@=11
\newcount\@tempcntc
\def\@citex[#1]#2{\if@filesw\immediate\write\@auxout{\string\citation{#2}}\fi
  \@tempcnta\z@\@tempcntb\m@ne\def\@citea{}\@cite{\@for\@citeb:=#2\do
    {\@ifundefined
       {b@\@citeb}{\@citeo\@tempcntb\m@ne\@citea\def\@citea{,}{\bf ?}\@warning
       {Citation `\@citeb' on page \thepage \space undefined}}%
    {\setbox\z@\hbox{\global\@tempcntc0\csname b@\@citeb\endcsname\relax}%
     \ifnum\@tempcntc=\z@ \@citeo\@tempcntb\m@ne
       \@citea\def\@citea{,}\hbox{\csname b@\@citeb\endcsname}%
     \else
      \advance\@tempcntb\@ne
      \ifnum\@tempcntb=\@tempcntc
      \else\advance\@tempcntb\m@ne\@citeo
      \@tempcnta\@tempcntc\@tempcntb\@tempcntc\fi\fi}}\@citeo}{#1}}
\def\@citeo{\ifnum\@tempcnta>\@tempcntb\else\@citea\def\@citea{,}%
  \ifnum\@tempcnta=\@tempcntb\the\@tempcnta\else
   {\advance\@tempcnta\@ne\ifnum\@tempcnta=\@tempcntb \else \def\@citea{--}\fi
    \advance\@tempcnta\m@ne\the\@tempcnta\@citea\the\@tempcntb}\fi\fi}
\catcode`@=12

\def\tanb{\tan\beta}
\def\mVV{M_{VV}}

\def\baselinestretch{1.25}

\begin{document}

\newlength{\captsize} \let\captsize=\small % use \let\normalsize=\captsize
%\newlength{\captwidth}                     % just before \caption{  ...

%\preprint{
%
\font\fortssbx=cmssbx10 scaled \magstep2
\hbox to \hsize{
%
%\special{psfile=uwlogo.ps
% hscale=8000 vscale=8000
% hoffset=-12 voffset=-2}
%\hskip.5in \raise.1in
%
$\vcenter{
\hbox{\fortssbx University of California - Davis}
%\hbox{\fortssbx University of Wisconsin - Madison}
}$
\hfill
$\vcenter{
\hbox{\bf UCD-97-17} 
\hbox{July 1997}
}$
}
%}

\vspace{.25in}

\def\mVV{M_{VV}}

%\begin{center}
%{\large\bf Muon Colliders: The Machine and The Physics}
%%\footnote{To appear in  {\it Proceedings of ``Beyond the Standard
%Model V'', Balholm, Norway, May, 1997}. Based on work performed in
%collaboration with V. Barger, M. Berger, and T. Han.} \\
%\vspace{.5cm}
%{\large John F. Gunion} \\
%\vspace{.1cm}
%{\sl Davis Institute for High Energy Physics}\\
%{\sl University of California at Davis, Davis, CA 95616, USA}\\
%\vspace{.5cm}
%\begin{abstract}
%\bigskip
%\baselineskip=13pt
%A very brief review of muon colliders is presented. Basic features
%of the accelerator and detector are outlined, and the very exciting
%physics prospects are reviewed.
%\end{abstract}
%\end{center}

\title{Muon Colliders: The Machine and The Physics
\footnote{To appear in  {\it Proceedings of ``Beyond the Standard
Model V'', Balholm, Norway, May, 1997}. Theoretical remarks are largely
based on work performed in collaboration with V. Barger, M. Berger, and T.
Han. The outline of the machine and detector is based on material provided by
R. Palmer and A. Tollestrup.} }
\author{John F. Gunion}  \address{Davis Institute for High Energy Physics,
Department of Physics\\ 
University of California at Davis, Davis, CA 95616, USA}
\maketitle
\thispagestyle{empty}

\begin{abstract}
A review of muon colliders is presented. Basic features      
of the accelerator and detector are outlined, and the very exciting
physics prospects are reviewed.
\end{abstract}

\section{Introduction}

\indent\indent
This review is divided into two sections.  In the first,
we outline basics of the muon accelerator complex and the detector,
noting critical requirements for optimal physics and the points
of greatest current concern and focus for future R\&D.
In the second section, the physics of the muon collider will
be high-lighted. On occasion,
I will note advantages, disadvantages and complementarity
relative to an $\epem$ collider. One finds that there
are important physics issues that
require both types of collider for 
the fullest and/or most precise results.

\baselineskip=14pt
\section{The Machine and Detector}

\indent\indent
The designs of the muon collider and associated detector have been
rapidly evolving in the last few 
years~\cite{mupmumi,saus,montauk,sanfran,feas,palgal}.
A muon collider (MC) facility can be developed in stages, each successive stage
building upon the previous one. Three stages are currently envisioned.
\begin{itemize}
\item Low-energy Higgs factory collider: $\rts\sim 100\gev$.
\item Intermediate-energy collider: $\rts\lsim 500\gev$.
\item High-energy collider: $\rts\sim 3-4\tev$.
\end{itemize}
The instantaneous luminosity, $\call$, that can be achieved at each stage 
is still somewhat uncertain. For rather conservative designs of relatively
low cost, current minimal expectations are:
\begin{itemize}
\item
$\call\sim 1,2,10 \times 10^{31}$cm$^{-2}$s$^{-1}$ at $\rts=100\gev$
for beam energy resolutions of $R=0.003\%,0.01\%,0.1\%$, respectively;
\item
$\call\sim 1 \times 10^{33}$cm$^{-2}$s$^{-1}$, at $\rts=300-500\gev$
for $R\sim 0.14\%$;
\item
$\call\sim 1 \times 10^{35}$cm$^{-2}$s$^{-1}$, at $\rts=3-4\tev$
with $R\sim 0.16\%$.
\end{itemize}
(For yearly integrated luminosities, we use the standard convention
of $\call=1\times 10^{32}$cm$^{-2}$s$^{-1}\To L=1\fbi/{\rm yr}$.)
It is believed that a combination of money and clever ideas
may allow the ultimate $\call$ values to be as much as
a factor of 5 to 10 larger than listed above. We shall occasionally
discuss the extent to which such higher luminosity is important for 
different types of physics.

The basic components of the collider are the following (see
Fig.~\ref{schematic}).
\begin{itemize}

\begin{figure}[p]
\epsfxsize=6.5in
\centerline{\epsffile{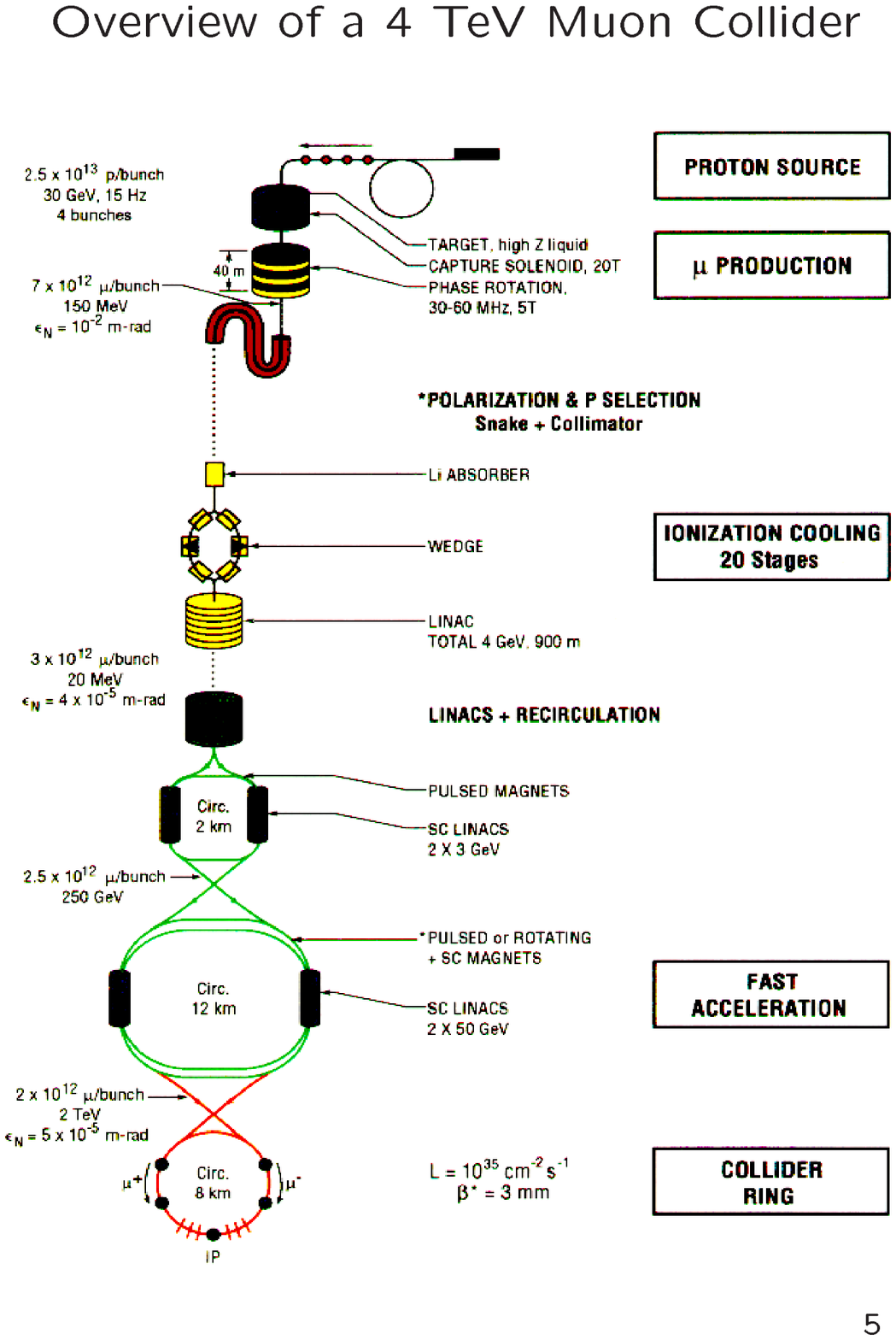}}
\vspace{10pt}
\caption[fake]{\baselineskip=0pt Muon Collider Schematic.}
\label{schematic}
\end{figure}

\item {\bf Proton Source:} 
One begins with a $\sim 600$ MeV Linac, feeding into a $\sim 3.6$ GeV Booster
(much like at BNL or SNS), which in turn feeds a $15-30$ GeV driver 
(much as envisioned for JHP \& Kaon). At lower (higher) energies, two (four)
bunches of $5\times 10^{13}$ ($2.5\times 10^{13}$) protons would be employed.
\item {\bf Target:} The goal is a large number of pions.
A good choice of target
might be liquid Ga. (It must be possible to cool the target
at these high intensities.)
\item {\bf Solenoid(s):} A high percentage of the produced low-energy
pions must be captured. A
$B=20$ T solenoid would be employed.  This would be followed by a 5 T
solenoidal decay channel in which the muons would emerge
and be retained.
\item {\bf Ionization Cooling:} The muons must next be cooled
very rapidly (given the finite muon lifetime) and with minimal
losses.  The following strategy is envisioned.  
\begin{itemize} 
\item One reduces both $p_T$ and $p_z$ using $dE/dz$ losses in,
\eg, Li.  
\item Next, the muons are accelerated to increase $p_z$, leaving $p_T$
unchanged. In this way, you are effectively cooling in $p_T$.
\item To cool in $p_z$, one introduces dispersion, i.e. separates
muons with different $p_z$, and then
uses a Be or Li wedge oriented so as to slow down large-$p_z$ muons relative
to small-$p_z$ muons, thereby cooling in $p_z$. The different
muon `streams' are then brought back together. 
\item This process is repeated many ($>20$) times.
\end{itemize}
\item {\bf Acceleration:} At higher energy, one possible (and possibly the
cheapest) approach is to employ synchrotrons with fast pulsed magnets
and long SC linacs. For 250 GeV beams, one would employ 4 T pulsed
magnets ($t=1$ msec). For 2 TeV beams, it would be necessary to
interlace fixed 8T SC dipole magnets with $\pm 2$ T pulsed magnets.
At Higgs factory energies, pulsed magnets would not be required
and recycling could be employed.
\item {\bf Collider Ring:} In order to maximize
the luminosity, the number of turns the muons make in the ring before
they decay should be as large as possible. The current plan
is for about 1000 turns, requiring a high field for the bending
magnets. Lattice designs involving octopole and quadrapole magnets
have been developed.
\end{itemize}

One critical issue for a muon collider is the nature
of the physics backgrounds. There are three major sources:
\begin{itemize}
\item {\bf The muon halo:} Muons lost from the main bunches can
still make it to the detector with full energy. These can
pass through the calorimeter and undergo deep-inelastic scattering.
It is found that this background can be adequately controlled by
careful injection and collimation.

\item {\bf Muon decay:} $\mu^-\to e^-\nu_\mu\anti\nu_e$ and $\mu^+\to
e^+\anti\nu_\mu\nu_e$ leads
to roughly $2\times 10^{12}\times 2$ muon decays during
the roughly 1000 turns during which a typical muon is stored
in the final ring. These decays give rise to two important effects:
(a) heating of the beam pipe, which must then be cooled, perhaps by a
water-cooled tungsten liner inside the magnets; (b) background
at the detector, which can be tamed by careful design of the
(Tungsten) nose cone at the detector entrance. 

\item {\bf Beam-beam interactions:}
Incoherent $\epem$ pair creation arises 
from beam-beam interactions at each crossing. The large
cross section ($\sigma\sim 10$
mb) yields about $3\times 10^4$ $\epem$ pairs per crossing.
Most of these soft $\epem$ pairs are curled up by the detector
solenoid magnetic field and collimated along the beam pipe.
The rest of the soft pairs are taken care of by a careful
design for the nose cone. Coherent pair creation has also been 
shown not to create a problem for the detector.

\end{itemize}
The current conclusion is that present state-of-the-art technologies
will be sufficient to build a detector 
able to handle the remaining background, consisting
primarily of a large number of soft particles, while achieving
normal standards for physics capabilities. 
The major current uncertainty is whether or not
the first layer of the vertex detector can be placed close enough
to the beam pipe to allow separation of charm from bottom quarks.

An issue currently being explored is whether or not the large number
of neutrinos that emanate from the storage ring pose a radiation
hazard as they interact over a period of time with the surrounding
earth and build up a low-level source of radiation.  The latest
calculations indicate that if the machine is built at a reasonable
depth, then this is
not a problem for center-of-mass energies up to about $3-4$ TeV.
For higher energies, wobbling of the beam orbit or some similar
technique could be employed.

At the time of this talk, the primary technical R\&D technical
developments that are needed in order to make a muon collider a
reality are: \begin{itemize}
\item demonstration of a working cooling system with small losses;
\item demonstration of the viability of low frequency linacs for phase
rotation and cooling;
\item development of the pulsed magnets, shielding
and SC r.f. cavities required for the accelerator at higher energies. 
\item demonstrated ability to construct the quadrapole magnets
required for the interaction region (expected to be easily achieved
for Higgs factory energies).
 \end{itemize}

I now list some of the important
$+$'s and $-$'s, as well as critical requirements and benchmarks for
the muon collider. 
First, there are some important advantages as compared to an electron
collider.
\begin{itemize} 
\item
There is less bremsstrahlung and no beamstrahlung.
\item
Beam energy resolution
can be substantially better --- in particular, with beam compression
techniques $R=0.003\%$ can be achieved at the low-energy Higgs factory
so that the Gaussian spread in $\rts$, given by
\begin{equation}
\sigrts\sim 2\mev\left({R\over 0.003\%}\right)\left({\rts\over
100\gev}\right)\,,
\end{equation}
can be as small as the natural width of a light SM-like Higgs boson.
\item The beam energy can be very precisely tuned: $\Delta E_{\rm beam}\sim
10^{-5}E_{\rm beam}$ is `easy'; $10^{-6}$ is achievable and very
important for scanning a narrow Higgs boson and precision $\mw$
and $\mt$ measurements. 
(To achieve such precision, power supplies stable at the $10^{-6}$
level are required and one must plan to monitor
the beam energy continuously via spin rotation measurements.)
\item
Multiple interaction regions in the final storage ring, allowing full
luminosity for several detectors, might not be impossible.
\end{itemize}
Other positive features of the muon collider include the following.
It can be built in stages.
The proton driver, intense muon source, cooled low-energy muon beam,
and so forth, that will sequentially  become available as the machine
is constructed would all have important uses of their own. The energy can be
increased by additions that are modest in physical
size and don't involve substantial
new technology. Particularly noteworthy are the following points.
\begin{itemize}
\item
If constructed at Fermilab,
the $\sim 50\gev$ $\mu^+$ and $\mu^-$ beams needed for the Higgs
factory could be collided with the $1\tev$ proton beam of the Tevatron,
yielding a $\mu p$ analogue of HERA with roughly $\sqrt 2$ times
as large center of mass energy and larger luminosity. Eventual higher
energy, higher luminosity muon beams would result in a $\mu p$
collider with physics reach vastly exceeding that of HERA.
\item Since the cost of a final storage ring is modest, several 
would be built as the energy of the machine is increased, each designed
to optimize luminosity at specific energies designed for
specific physics goals (to be discussed in more detail later).
An incomplete list is the following.
\begin{itemize}
\item  
If a light ($\mh\lsim 2\mw$) 
SM-like Higgs boson has been observed (\eg\ at the LHC), 
the first energy goal and ring constructed would
be for factory-like $s$-channel production and study
at $\rts\sim\mh$~\cite{higgsi,schannelreport}. 
\item
A second energy goal and ring would be for
operation at high $\call$ near the $Z\h$ threshold. 
(This would actually be the first goal if a SM-like Higgs
has been observed and has $\mh>2\mw$.) One would
choose $\rts$ so that  the $Z\h$ cross section
is maximal, thereby allowing precise measurement of many
Higgs boson properties. (Even if $\mh<2\mw$, there are important
Higgs properties that are not easily measured in $s$-channel
production.) A fairly precise determination of
$\mh$ from the $\sigma(Z\h)$ threshold
rise would also be possible~\cite{zhthreshold}.
\item
Exceptionally precise measurements of
$\mw$ and of $\mt$, $\alpha_s$, $\Gamma_t$, are possible~\cite{mwmt}
with rings that achieve full luminosity at
$\rts\sim 2\mw$ and/or $\rts\sim 2\mt$,  respectively.
If no Higgs boson is seen at the LHC, then this would constitute
an important first goal for the muon collider.
\item 
Factories for $s$-channel production of any new particle
with $\mupmum$ couplings would be possible. Possibilities
include a new $Z^\prime$  and a sneutrino 
with R-parity-violating coupling to $\mupmum$.
\end{itemize}
Once the accelerator is operating at high energy, 
beams of different energy appropriate to the
different rings could be extracted and the luminosity could be shared
among the various rings (and with the $\mu p$ collider).
This would allow simultaneous pursuit of many different types of physics
at different detectors, as possibly desirable from both a physics 
and a sociological point of view.
\end{itemize}
There are two clear disadvantages of a muon collider:
\begin{itemize}
\item A $\gam\gam$ collider is not possible at a muon collider facility.
\item Some polarization is automatic, but
large polarization implies sacrifice in luminosity at a muon collider.
This is because large polarization is achieved by keeping
only the larger $p_z$ muons emerging from the target, rather than collecting
nearly all the muons.
\end{itemize}

\section{The Physics}

Early studies~\cite{sausth,sanfranth}
made it clear that a muon collider would be an extremely
valuable tool for exploring the physics of any
conceivable extension of the Standard Model (SM). Fully detailed
studies are now available for most types of new physics.
To illustrate the results, I shall briefly discuss:
\begin{itemize}
\item Higgs physics.
\item Strong $WW$ sector physics.
\item A new $Z^\prime$.
\item Precision $\mw$ and $\mt$ measurements.
\item Standard supersymmetry.
\item R-parity violation phenomena in supersymmetry.
\item Leptoquarks.
\end{itemize}

\subsection{Higgs Physics}

If the $\mupmum$ collider is operated 
by running at the highest energy or at the maximum in the $Z\h$ 
cross section, then it will have similar capabilities to an $\epem$
collider operating at the same $\rts$ and $\call$ 
(barring unexpected detector backgrounds at the muon collider).
The totally unique feature of a muon collider is the possibility
of  $s$-channel Higgs production, $\mupmum\to \h$, which can
have a very high rate if the total Higgs width, $\gamh$, and the
beam energy resolution, $R$, are both small.
The importance of small $R$ and small $\gamh$ is evident from 
the result, $\sighbar$, of convoluting a Gaussian $\rts$ distribution of width
$\sigrts$ with the standard $s$-channel Breit Wigner Higgs resonance 
cross section.  For $\rts=\mh$, one obtains
\begin{equation}
\sighbar\simeq {\pi\sqrt{2\pi} \Gamma(\h\to\mu\mu)\, \br(\h\to X) \over
\mh^2\sigrts}
\times \left(1+ {\pi\over 8}\left[{\gamh\over
\sigrts}\right]^2\right)^{-1/2}\,. 
\label{sigmaform}
\end{equation}
Eq.~(\ref{sigmaform}) shows that
the smaller $\gamh$ is, and the more nearly $\sigrts$ can
be made comparable to $\gamh$, the larger will be $\sighbar$.
Although smaller $R$ implies smaller $\call$, one finds
that for a Higgs boson with a very narrow width, \eg\ a SM-like Higgs boson
with $\mh\lsim 2\mw$,  
it is advantageous to use the smallest $R$ that can be achieved.
A Higgs boson with a large width
will only be visible at a muon collider if its
$\mupmum$ coupling (and, hence, partial width) is enhanced relative
to that of a SM Higgs boson.

Below, we update results obtained (assuming $R=0.01\%$
and $L\sim 50\fbi/{\rm yr}$) in 
Refs.~\cite{higgsi}, \cite{schannelreport}
and \cite{snowmasssummary} to account for
the preliminary Higgs-factory design parameters resulting from the 
recent detailed study of the low-energy machine --- we employ $R=0.003\%$
and compare luminosities of $L=0.1\fbi/{\rm yr}$ and $1\fbi/{\rm yr}$,
the former being conservative and the latter optimistic. With regard to
the statistical accuracy for various measurements,
compared to the earlier studies the lower expected $L$ is only partially
offset by the smaller $R$. However, since $\sigrts\sim \gamhsm$
for $R=0.003\%$, systematics in measuring $\gamh$ (via scanning) 
associated with imperfect knowledge of the exact shape of the $\rts$
spectrum (in particular its wings) are much less of a concern than
for $R=0.01\%$.

\subsubsection{A Standard Model-Like Higgs Boson}

In all likelihood, the $\h$ will already have been discovered
at either the LHC or NLC, if not at LEP or the Tevatron, by the time the muon
collider is built, and its mass will have been accurately measured:
$\Delta\mh\sim 100\mev$ for $L=300\fbi$ at the LHC; $\Delta\mh\sim 50\mev$
for $L=200\fbi$ in $\rts=500\gev$ running at the NLC
\cite{snowmasssummary}. If $\mh>2\mw$, $\gamh$ will be large, leading to tiny 
$\br(\h\to\mupmum)$; the resulting $\sighbar$ is too small to be seen
above background in $s$-channel production
at a muon collider. If $\mh <2\mw$, $\sighbar$ will be large and
a Higgs-factory muon collider ring 
with optimal luminosity at $\rts\sim\mh$ will be a high priority. 
At the muon collider,
the first task will be to scan over the $\Delta\mh$ interval so as
to center on $\rts\simeq\mh$ within a fraction of $\sigrts$.
A ``typical case'' is $\mh\sim 110\gev$, $\sigrts\sim 2\mev$,
$\Delta\mh\sim 100\mev$. About  $50$ scan points are needed to center on
$\rts\simeq\mh$ within $0.3 \sigrts$. Each point requires $L\sim 0.0015\fbi$ to
observe or eliminate the $\h$ at the $3\sigma$ level. A total of up to
$L=0.075\fbi$ would then be needed for the centering process. Thus, for
$L=0.1\fbi/{\rm yr}$ centering might take the better part of a year.
The worst case is $\mh\sim \mz$ --- with $\sigrts\sim 2\mev$ and $\Delta\mh\sim
100\mev$, up to a factor of $50$ more $L$ 
would have to be devoted to the centering process; even for $L=1\fbi/{\rm yr}$
the nearly four years required would be unacceptable.

Once we are able to center on the Higgs peak, the measurements of primary
importance are the Higgs total width and the
cross sections $\sigma(\mupmum\to \h\to X)$ for $X=b\anti
b,W\wstar,Z\zstar$.\footnote{Note from Eq.~(\ref{sigmaform})
that $\sigma(\mupmum\to\h\to X)$ provides
a determination of $\Gamma(\h\to\mupmum)\br(\h\to X)$ unless $\sigrts\ll
\gamh$.}  To measure all of these simultaneously, it is best
to employ an optimized three-point scan of the 
Higgs peak \cite{schannelreport}. 
The accuracies of the measurements for total
luminosities of $L=4\fbi$ and $0.4\fbi$ (four year's of running
at $L=1\fbi/{\rm yr}$ and $0.1\fbi/{\rm yr}$, respectively) are
tabulated in Table~\ref{fmcerrors}. Note that at $L=0.4\fbi$ 
the errors for $\sigma\br(\hsm\to b\anti b)$ are still
generally small but that those for $\gamhsm$ are uncomfortably large. In fact,
errors for $\gamhsm$ obtained
indirectly using a combination of $L=600\fbi$ LHC data, $L=200\fbi$ NLC data,
and $L=50\fbi$ $\gam\gam$-collider data are often better: $\sim 19\%$ 
for $\mhsm\lsim 120\gev$ and $\sim 10\%-13\%$ for $130\gev\lsim \mhsm\lsim
180\gev$.

\begin{table}[h]
\caption[fake]{\baselineskip 0pt Percentage errors ($1\sigma$) for 
$\sigma\br(\hsm\to b\anti b,W\wstar,Z\zstar)$
(extracted from channel rates)
and $\gamhsm$ for $s$-channel Higgs production at the MC
assuming beam energy resolution of $R=0.003\%$. Results are
presented for two integrated four-year luminosities:
$L=4\fbi$ ($L=0.4\fbi$). An optimized three-point scan
is employed [which, for the cross section measurements, is
equivalent to $L\sim 2\fbi$  ($L=0.2\fbi$) at the $\rts=\mhsm$ peak].
It is useful to compare this table to the $L=200\fbi$, $R=0.01\%$ table
of Ref.~\protect\cite{snowmasssummary}.}
\small
\begin{center}
\begin{tabular}{|c|c|c|c|c|}
\hline
 Quantity & \multicolumn{4}{c|}{Errors} \\
\hline
\hline
 {\bf Mass (GeV)} & {\bf 80} & {\bf $\mz$} & {\bf 100} & {\bf 110} \\
\hline
$\sigma\br(b\anti b)$ & $0.8\%(2.4\%)$ & $7\%(21\%)$ & $ 1.3\%(4\%)$ & 
$ 1\%(3\%)$ \\
\hline
$\sigma\br(W\wstar)$ & $-$ & $-$ & $ 10\%(32\%)$ & 
$ 5\%(15\%)$ \\
\hline
$\sigma\br(Z\zstar)$ & $-$ & $-$ & $-$ & 
$ 62\%(190\%)$ \\
\hline
 $\gamhsm$ & $ 3\%(10\%)$ & $ 25\%(78\%)$ & $ 10\%(30\%)$ & 
  $ 5\%(16\%)$ \\
\hline
\hline
 {\bf Mass (GeV)} & {\bf 120} & {\bf 130} & {\bf 140} & {\bf 150} \\
\hline
$\sigma\br(b\anti b)$ & $1\%(3\%)$ & $1.5\%(5\%)$ & $ 3\%(9\%)$ & 
$ 9\%(28\%)$ \\
\hline
$\sigma\br(W\wstar)$ & $3\%(10\%)$ & $2.5\%(8\%)$ & $ 2.3\%(7\%)$ & 
$ 3\%(9\%)$ \\
\hline
$\sigma\br(Z\zstar)$ & $16\%(50\%)$ & $10\%(30\%)$ & $ 8\%(26\%)$ & 
$ 11\%(34\%)$ \\
\hline
 $\gamhsm$ & $ 5\%(16\%)$ & $ 6\%(18\%)$ & $ 9\%(29\%)$ &
  $ 34\%(105\%)$ \\
\hline
\end{tabular}
\end{center}
\label{fmcerrors}
\end{table}

The $s$-channel measurements can then be combined with LHC data and 
data from NLC (or MC) running at $\rts>\mz+\mh$ in order to determine all the
properties of the $\h$ in a model-independent way.
For example, there will be four ways to determine
$\Gamma(\h\to\mupmum)$: 
\begin{eqnarray}
{ 1)}~~~  \Gamma(\h\to\mupmum)&=&{[\Gamma(\h\to\mupmum)\br(\h\to
b\anti b)]_{\rm MC}\over \br(\h\to b\anti b)_{\rm NLC}} ;\nonumber\cr
{ 2)}~~~  \Gamma(\h\to\mupmum)&=&{[\Gamma(\h\to\mupmum)\br(\h\to
W\wstar)]_{\rm MC}\over \br(\h\to W\wstar)_{\rm NLC}} ;\nonumber\cr
{ 3)}~~~  \Gamma(\h\to\mupmum)&=&{[\Gamma(\h\to\mupmum)\br(\h\to
Z\zstar)]_{\rm MC}\gamh\over \Gamma(\h\to Z\zstar)_{\rm NLC}} ;\nonumber\cr
{ 4)}~~~  \Gamma(\h\to\mupmum)&=&{[\Gamma(\h\to\mupmum)\br(\h\to
W\wstar)\gamh]_{\rm MC}\over \Gamma(\h\to W\wstar)_{\rm NLC}} .\nonumber
\end{eqnarray}
The associated errors for the SM Higgs are labelled $(\mupmum\hsm)^2|_{\rm
NLC+MC}$ in Table~\ref{nlcfmcerrors} below.

\begin{table}[h]
\caption[fake]{Percentage errors ($1\sigma$)
for combining $L=200\fbi$--$\rts=500\gev$ NLC, $L=600\fbi$ LHC, 
$L=50\fbi$ $\gam\gam$-collider 
and MC $R=0.003\%$ $s$-channel data, with errors for the latter as quoted
in Table~\ref{fmcerrors}. Results are presented for two total four-year
integrated MC luminosities: $L=4\fbi$ ($L=0.4\fbi$)
Comparison to the similar table of Ref.~\cite{snowmasssummary},
which assumed $L=200\fbi$ for $R=0.01\%$ at the MC, is useful.}
\small
\begin{center}
\begin{tabular}{|c|c|c|c|c|}
\hline
 Quantity & \multicolumn{4}{c|}{Errors} \\
\hline
\hline
 {\bf Mass (GeV)} & {\bf 80} & {\bf 100} & {\bf 110} & {\bf 120} \\
\hline
 $(b\anti b\hsm)^2|_{\rm NLC+MC} $ & $6\%(10\%)$ & $ 10\%(16\%)$ & 
$ 7\%(13\%)$ &   $7\%(13\%)$ \\
\hline
 $(c\anti c\hsm)^2|_{\rm NLC+MC} $ & $9\%(13\%)$ & $ 12\%(18\%)$ & 
$ 10\%(15\%)$ &   $10\%(15\%)$ \\
\hline
 $(\mupmum\hsm)^2|_{\rm NLC+MC}$ & 
$ 5\%(5\%)$ & $ 5\%(5\%)$ & $ 4\%(5\%)$ & $ 4\%(4\%)$ \\
\hline
 $(\gam\gam\hsm)^2|_{\rm MC}$ & $ 15\%(18\%)$ & $ 17\%(33\%)$ & 
$ 14\%(21\%)$ &  $ 14\%(20\%)$ \\
\hline
 $(\gam\gam\hsm)^2|_{\rm NLC+MC}$ & $ 9\%(10\%)$ & $ 10\%(11\%)$ & 
$ 9\%(10\%)$ &  $ 9\%(10\%)$ \\
\hline
\hline
 {\bf Mass (GeV)} & {\bf 130} & {\bf 140} & {\bf 150} & {\bf 170} \\
\hline
 $(b\anti b\hsm)^2|_{\rm NLC+MC}$ & $8\%(12\%)$ & $9\%(10\%)$ & 
$13\%(13\%)$ &  $23\%(23\%)$ \\
\hline
 $(c\anti c\hsm)^2|_{\rm NLC+MC} $ & $10\%(14\%)$ & 
\multicolumn{3}{c|}{$?$} \\
\hline
 $(\mupmum\hsm)^2|_{\rm NLC+MC}$ & 
$ 4\%(5\%)$ & $ 4\%$ & $ 4\%(5\%)$ &  $ 13\%(14\%)$ \\
\hline
 $(W\wstar\hsm)^2|_{\rm MC}$ & $ 17\%(24\%)$ & $ 12\%(30\%)$ & 
$ 33\%(104\%)$ &  $-$ \\
\hline
 $(W\wstar\hsm)^2|_{\rm NLC+MC}$ & $ 5\%$ & $ 5\%$ & $ 6\%(8\%)$ &
 $ 10\%$ \\
\hline
 $(\gam\gam\hsm)^2|_{\rm MC}$ & $ 14\%(22\%)$ & $ 20\%(34\%)$ & 
$ 48\%(110\%)$ & $-$ \\
\hline
 $(\gam\gam\hsm)^2|_{\rm NLC+MC}$ & 
    $ 10\%(12\%)$ & $ 13\%(15\%)$ & $ 25\%(29\%)$ & $-$ \\
\hline
\end{tabular}
\end{center}
\label{nlcfmcerrors}
\end{table}

Errors as small as given in Table~\ref{nlcfmcerrors}
may make it possible to distinguish between 
the SM $\hsm$ and the $\hl$ of the MSSM~\cite{snowmasssummary}. 
If deviations from SM predictions
are apparent, then an approximate determination of the crucial
MSSM CP-odd Higgs boson mass $\mha$ can be made.
(The $\hl$ becomes indistinguishable from the $\hsm$ if $\mha$ is large ---
the decoupling limit.) The most useful quantity for this purpose 
if only $s$-channel Higgs factory 
MC data are available (\ie\ no $Z\h$ NLC or MC
data) is the coupling-squared
ratio $(W\wstar\hsm)^2/(b\anti b\hsm)^2\propto \sigma\br(W\wstar)/
\sigma\br(b\anti b)$. 
If $110\lsim\mhsm\lsim 140\gev$ (a very likely region in the MSSM)
then this ratio will
be measured with a statistical accuracy of $\lsim \pm
5\%$ for $L=4\fbi$ (see Table~\ref{fmcerrors}). Systematic 
errors of order $\pm 5\%-\pm10\%$
from uncertainty in the $b$ quark mass will also enter
the interpretation of this ratio. A $>2-3\sigma$
deviation will be observed if $\mha<450\gev$. For $L=0.4\fbi$, one
would observe a $>1.5-2\sigma$ deviation for $\mha<450\gev$.
If $Z\h$ data from the NLC (or MC) {\it is} available then the best
quantity for discriminating between the $\hl$ and $\hsm$ 
is the fundamental coupling $\Gamma(\h\to\mupmum)$.
For all $\mh\lsim 2\mw$, the error in
$\Gamma(\h\to\mupmum)$ obtained after combining NLC and MC data 
as sketched in the equations above
is dominated by the NLC denominators 
and (for $L=200\fbi$ at the NLC) is $\lsim 5\%$, even for $L=0.4\fbi$ (see
Table~\ref{nlcfmcerrors}). This
will allow detection of a $>3\sigma$ deviation from
the SM value if $\mha<600\gev$. Systematic errors from theoretical
uncertainties in the interpretation of this measurement are small.
Note that $\gamh$ alone cannot be used to distinguish between the 
MSSM and SM in a model-independent way. This is  because
$\gamh$ depends on
many things, including (in the MSSM) the squark-mixing model.

\subsubsection{MSSM $\hh$ and $\ha$}

One of the potentially most important features of a muon collider is that
the $s$-channel processes $\mupmum\to\hh,\ha$ allow production
and study of $\hh,\ha$ up to $\mha\sim\mhh\lsim \rts$~\cite{schannelreport}. 
Discovery possibilities at other colliders are more limited (a 
more detailed summary
and references appear in~\cite{snowmasssummary}): (a) at the LHC,
discovery of $\hh,\ha$ is not possible for $\mha\gsim 200\gev$ at moderate
$\tanb\gsim 3$; (b) at $\rts=500\gev$, $\epem\to \hh\ha$ 
pair production probes only to $\mha\sim\mhh\lsim 230-240\gev$; (c) 
a $\gam\gam$ collider could potentially probe up
to $\mha\sim\mhh\sim 0.8 \rts\sim 400\gev$ with $L\gsim 150-200\fbi$.
The reach of the muon collider depends very much on the
$\tanb$ parameter of the MSSM and the luminosity
achievable at intermediate energies.
A total of $L=200\fbi$ must be used in scanning the $200\leq\rts \leq 500\gev$
interval to guarantee discovery
of the $\hh,\ha$ for the $\tanb\geq 3$ portion of parameter
space such that the $\hh,\ha$ cannot be discovered at the LHC
in this same mass interval. The conservative intermediate-energy-collider
luminosity corresponds
to $L\sim 40\fbi$ over four years, for which one can only
reach down to $\tanb\geq 5-6$. 
(If $\tanb$ is still larger and the MC is run at $\rts=500\gev$, the $\hh,\ha$
can also be discovered in the bremsstrahlung tail if the $b\anti b$
mass resolution is good enough.) Once discovered,
the $\hh,\ha$ can be studied with precision at the $\mupmum$ collider.
In particular, only a direct $s$-channel scan may allow separation of
the $\hh$ from the $\ha$ when they are approximately degenerate (as
predicted for large $\tanb$).

Even masses as large as  $\mha\sim\mhh\sim\mhpm> 1\tev$ 
cannot be ruled out simply on the basis of hierarchy/naturalness, although
the model would be fine-tuned. Discovery of the $\hh,\ha,\hpm$ via
$\epem\to\hh\ha,\hp\hm$ would require $\rts_{\epem}>2\tev$, 
currently thought difficult
to achieve. In contrast, it is currently expected that a muon collider with
$\rts\sim 3-4\tev$ is feasible, in which case $\mupmum\to \hh\ha,\hp\hm$
observation would be straightforward. Studies~\cite{kg,fengmoroi}
show that the $\hh,\ha$ could be
detected in their $b\anti b$ or $t\anti t$  decay modes and $\hpm$ in 
$t\anti b$ and $b\anti t$ decays, even if SUSY decays are present.
Measurements of relative branching
ratios for $\hh,\ha,\hpm$ decays to different final states
(including SUSY channels)
can also be performed with good accuracy. The branching ratio results,
together with the determination of $\mha\sim\mhh\sim\mhpm$ and, say, $\mcpmone$
(the light chargino mass), allow one to discriminate with incredible
statistical significance between different closely similar GUT
scenarios~\cite{kg}.

\subsubsection{Exotic Higgs Bosons}

A muon collider could play a very important role in probing an exotic
Higgs sector, even if the exotic Higgs bosons have already been detected 
at another accelerator. To give one example, consider a Higgs sector
containing a doubly-charged Higgs boson, $\dmm$, that is a member
of a SU(2)$\times$U(1) representation that either has no neutral member
or a neutral member with zero vacuum expectation value (as required
for $\rho\equiv\mw/[\mz\cos\theta_W]=1$ to be natural).
For many choices of
representation, $e^-e^-\to \dmm$ and $\mu^-\mu^-\to\dmm$ couplings are allowed.
(We denote the Majorana-like coupling strengths by $\lam_{ee,\mu\mu}$.)
A $\dmm$ with $\mdmm<500-900\gev$ (depending upon dominant decay)
will be seen previously at the LHC, if not TeV33~\cite{kpitts}. 
Once $\mdmm$ is known, observation of the $s$-channel processes
$e^-e^-\to\dmm$ and $\mu^-\mu^-\to\dmm$ will be possible and probably
would be the only means of directly measuring $\lam_{ee,\mu\mu}$~\cite{gdmm}.
For couplings not too far below current bounds, factory-like production rates
are predicted for the $\dmm$. For very small couplings
(such as those that might be associated with left-right symmetric models) the
very excellent $R=0.003\%$ beam energy resolution that can be achieved at a  
$\mu^-\mu^-$ collider implies that it can probe $\lam_{\mu\mu}$ magnitudes
that are significantly smaller than the $\lam_{ee}$
values that can be probed in $e^-e^-$ collisions. 

\subsection{Strong {\boldmath $WW$} Interactions and Related Models}

If no light SM-like Higgs is found at the LHC, NLC or MC, then
signals of the concomitant strongly-interacting $WW$ sector will be
found~\cite{wwlowlhc,wwlownlc}. 
However, as detailed in Ref.~\cite{wwhigh}, 
to fully explore strong $WW$ interactions requires
quark, electron or muon collision energies of $\rts\geq 3-4\tev$,
with appropriately matched luminosity. Such energies may be
most easily achievable at a muon collider. A muon collider with this
energy could study (using both $\mupmum$ and $\mu^-\mu^-$ collisions)
all isospin channels. In close
analogy to $\pi\pi$ scattering studies, 
different models could be distinguished from one another by
the detailed $WW$ mass spectra in the different isospin channels.
After several years of running at design luminosity, statistics would
be such that one could separately project out the  cross sections for
different final state polarizations, $W_LW_L$, $W_LW_T$ and $W_TW_T$.
Only by such precision studies would it be possible to determine in
detail the effective `chiral' Lagrangian for the strong $WW$ sector.

\subsection{A New {\boldmath $Z^\prime$}}

At a high energy $\mupmum$ collider, a new $Z^\prime$ with
$m_{Z^\prime}\leq\rts$ is easily discovered in the bremsstrahlung
tail of the $\mupmum$ energy spectrum.
Once found, a typical $Z^\prime$ would be produced with factory-like
rates if a specialized storage ring for $\rts\simeq m_{Z^\prime}$ is built.
(See Ref.~\cite{sf95} for details.)
The machine energy could either be set to this $\rts$, or
muons of appropriate energy could be extracted early 
in the acceleration process if the machine is run at higher energy. 

\subsection{Precision Measurements of {\boldmath $\mw$} and {\boldmath $\mt$}}

The comparison of electron and muon colliders for such measurements has been
studied in Ref.~\cite{mwmt} (see also \cite{dawson}).
At the NLC, $\mw$ is best determined via $q\anti q$ mass reconstruction
at $\rts=500\gev$~\cite{nlcmw} (see also \cite{lepii}) 
and $\mt$ via $t\anti t$ threshold measurements~\cite{nlcmt}.
The resulting precisions are \begin{equation}
\Delta \mw = 20 {\rm\ MeV}, \qquad
\Delta \mt = 0.2\rm\ GeV  \qquad (50\rm\ fb^{-1},\ NLC) \,.\label{nlc}
\end{equation}
Systematic effects deriving from beam energy spread and beam energy
uncertainty are such that the $\mw$ precision could not be
improved by running at the $WW$ threshold.
At the MC, the one part per million accuracy for the beam energy
and the small beam energy spread imply greater precision
for the $WW$ threshold and $t\anti t $ threshold measurements.
For $R\lsim 0.1\%$, 
\begin{equation} \Delta \mw = 9 {\rm\ MeV}, \qquad
\Delta \mt = 0.1\rm\ GeV  \qquad (50\rm\ fb^{-1},\ MC) \,.\label{fmc}
\end{equation}
where systematic effects have been included. To achieve the
indicated $\mw$ precision, the relative luminosity for $\rts=161\gev$
and $\rts=150\gev$ measurements would need to be well measured.
Even for $L=100\fbi$, errors would probably still be statistics
dominated at a $\mupmum$ collider, in which case one could achieve
\begin{equation}
\Delta \mw = 6 {\rm\ MeV}, \qquad
\Delta \mt = 0.07\rm\ GeV  \qquad (100\rm\ fb^{-1},\ MC) \,.\label{fmc2}
\end{equation}
Relatively modest improvements in the conservative
$\call$ expectations for $R\sim 0.1\%$ muon collider designs
(see introduction) would allow $L=50-100\fbi$ to be accumulated after
$\lsim 5$ years at $\rts\sim 2\mt$; more substantial improvements
in $\call$ expectations at $\rts\sim 2\mw$ would be needed.

\subsection{Standard SUSY Studies}

If R-parity is conserved, supersymmetric particles must be produced
in pairs at a lepton collider, requiring center-of-mass energy
greater than the sum of the masses. Although fine-tuning
considerations suggest that the lightest gauginos should have
$\mchitil\lsim 200-400\gev$, it is entirely possible for sfermions,
especially the squarks of the first and second generation, to
have masses $\gsim 1\tev$ without violating either
fine-tuning or considerations of naturalness/hierarchy. Further, gauge
unification is most successful if there are SUSY particles above
$1\tev$. The LHC will set the mass scale of the squarks, but will not
be able to determine their masses and decays in much detail
if they are heavy (due to limited event rates after cuts
required to control backgrounds). To study sfermions with mass of order
$1\tev$, an $\epem$ or $\mupmum$ collider would need $\rts\gsim
2.5\gev$ --- the $\beta^3$ p-wave threshold behavior for scalar
pair production implying a slow rise in the pair cross section
above threshold. The $\rts=3-4\tev$ option discussed as a possibility
for a $\mupmum$ collider would imply pair production
rates (for planned luminosity) adequate for detailed studies of the
sfermions~\cite{susystudies}.

\subsection{Leptoquarks and R-parity Violating SUSY Scenarios}

The HERA event excess at high-$Q^2$ with a possible resonance
component at $M_{e^+q}\sim 200\gev$ has led to a resurgence
of popularity for models with leptoquarks, including SUSY models
in which squarks play the role of leptoquarks. In the
latter case, the $\ell q\to\sq$
coupling  derives from R-parity violating Yukawa-like superpotential terms.
If the resonance signal holds up with increased statistics, then
it will be of great importance to search for a large variety of
closely related signals.  Of particular importance will be the question
of the flavor structure of leptoquarks, in particular whether
there  are $\mu q$
leptoquarks as well as $e q$ leptoquarks.  A natural way to explore
for the former is via $\mu^{\pm} p$ collisions at high luminosity, 
as possible at a muon collider facility by colliding one of the muon 
beams with protons of sufficient energy. 

Let us~\cite{mup} compare $e p$ collisions at HERA ($\rts\sim 314\gev$)
to $\mu p$ collisions
of the Higgs-factory $50\gev$ muon beams with the $1\tev$ Fermilab
Tevatron beam at the Main Injector ($\rts\sim 447\gev$). 
If the muon beam is extracted with $R\sim 0.1\%$ (\ie\ before
compression to $R=0.003\%$ or with compression turned off), then
yearly luminosity of $L\sim 1\fbi/{\rm yr}$ would be possible,
as compared to the $L\sim 0.1\fbi/{\rm yr}$ luminosity for HERA.
At HERA the `observed' leptoquark resonance probably contains
the proton valence quarks, \ie\ is of either
the $LQ=ed$ or $eu$ type.  To avoid flavor-changing neutral currents,
the muon-type leptoquarks are most naturally chosen to be 
the 2nd family $\mu s$ and $\mu c$ analogues. Let us denote
the $LQ \to \ell q$ coupling as $\lambda_{\ell q}^J$, where $J$
is the spin of the leptoquark. For scalar and vector
leptoquarks with mass $M_{LQ}=200\gev$, and assuming $\br(LQ \to \ell q)=1$,
5 LQ events are predicted
at HERA with $L=0.1\fbi$ for: $\lambda^0_{eu}=0.006$,
$\lambda^0_{ed}=0.012$, $\lambda^1_{eu}=0.004$, and $\lambda^1_{ed}=0.008$.
To normalize,
the observed HERA excess corresponds to $\lambda^0_{e^+d}\sim 0.025$.
At this same $M_{LQ}=200\gev$, the Higgs-factory/MI $\mu p$ collider
with $L=1\fbi$ yields 5 LQ events for: $\lambda^0_{\mu c}=0.007$,
$\lambda^0_{\mu s}=0.006$, $\lambda^1_{\mu c}=0.005$, 
and $\lambda^1_{\mu s}=0.004$. Given that 2nd family
leptoquark couplings will probably be larger than 1st family couplings, 
the Higgs-factory/MI $\mu p$ collider would be a very important facility 
if leptoquarks exist.

If the leptoquarks turn out to be squarks, then it will be important
to ascertain the complete structure of the R-parity
violating superpotential. The most general superpotential that 
violates lepton number while conserving baryon
number (in order to ensure proton stability) takes the form
\beq
W=\lam_{ijk}\what L_L^i\what L_L^j \what{\anti E_R^k}
+\lam^\prime_{ijk} \what L_L^i\what Q_L^j \what{\anti D_R^k}\,.
\eeq
The $\lam^\prime$ type interactions would 
lead to squark production in $e^+d$ collisions at HERA.
If a non-zero value for some of the $\lam^\prime$'s is confirmed,
it is very possible that one or more of the $\lam$'s is
also non-zero. Resonant $s$-channel sneutrino production --- $\epem\to
\snu_\tau$ ($\lam_{131}$), $\epem\to \snu_\mu$ ($\lam_{121}$),
$\mupmum\to \snu_\tau$ ($\lam_{232}$), and $\mupmum\to\snu_e$ ($\lam_{212}$)
--- would provide a particularly sensitive probe. 
To give one sample number, suppose $\lam=0.01$, $\msnu=100\gev$, and
$\mcnone=90\gev$. For these choices, $\Gamma^{\rm
tot}_{\snu}=0.52\gev$, including $\snu_\ell\to \nu_\ell \cnone$ decays. The
$\mu^+\mu^-\to\snu\to\mu^+\mu^-$ rate, $S$, for the resonance signal is
computed by convoluting a standard $s$-channel resonance form with 
the luminosity distribution as a function of $\rts$.  The latter is obtained
by assuming a Gaussian distribution in $\rts$, modified by the effects of
initial state bremsstrahlung from the incoming muons. The maximum $S$ is
obtained when the Gaussian distribution is centered at $\rts=\msnu$.  The same
$\rts$ distribution is used to compute the continuum background, $B$. The
resulting statistical significance of the signal is defined as $N_{\rm
SD}=S/\sqrt B$. Assuming  a beam energy resolution of $R=0.1\%$
and adopting the associated (conservative) 
integrated luminosity at $\rts\sim 100\gev$ of $L=1\fbi$,
one finds $S=1.7\cdot 10^3$ and $N_{\rm
SD}\sim 8$. At fixed $L$, 
$N_{\rm SD}$ decreases for larger $R$, and for the same underlying
$R$, is smaller for the equivalent $\epem\to\snu\to\epem$ situation because of
the increased bremsstrahlung and non-negligible beamstrahlung.
The typical S-band $\epem$ collider design has significant beamstrahlung 
and underlying beam energy resolution of $R\sim 1\%$. In combination,
the result is roughly the same as a beam energy resolution of $R\sim
3\%$. For the above parameter choices, one obtains $S=70$ and $N_{\rm SD}=0.3$.
Thus, higher luminosity is required to probe the same coupling level in the
$\epem$ case. The most important point is that both a muon collider and an
electron collider would be required in order to explore the flavor structure
of the $\lam$'s as fully as possible. 

\section{Conclusion}

The physics motivation for a muon collider is very strong. Different
types of physics would be probed, both as the collider complex is constructed,
and, once fully operational, as the energy of the muon beams is increased.
Complementarity to other planned and existing facilities would be enormous:
\begin{itemize}
\item
If the $ep$ HERA leptoquark signal persists,
the $\mu p$ collider that would be a natural spin-off
at a muon collider facility would be mandatory.
\item
Both a muon collider and an electron collider are needed to
understand the flavor structure of new physics in lepton-lepton channels.
\item
An $\epem$ collider focusing on $Z\h$ production in combination
with a $\mupmum$-collider Higgs factory will allow us
to fully explore the properties of a light SM-like Higgs boson
in the shortest time.
\end{itemize}
In addition, the muon collider would have unique capabilities. For example:
\begin{itemize}
\item
It would have the ability to observe the MSSM heavy Higgs bosons up
to the maximum $\rts$ available, using $s$-channel production.
\item
If there is new physics at high $\rts$ (supersymmetry, contact
interactions, $\ldots$) then a muon collider would be
critical if the necessary center-of-mass 
energy can only be economically achieved in muon collisions. 
\end{itemize}
Studies of, and R\&D for, a muon collider should be vigorously pursued.

\section{Acknowledgements}
This work was supported in part by the Department of Energy and by
the Davis Institute for High Energy Physics.
I am grateful to J. Gallardo, S. Geer, B. Palmer and A. Tollestrup
for helpful discussions and comments.


\begin{references}
\bibitem{susystudies} Basic results appear in Ref.~\cite{sf95}.
Further detailed studies are in progress.
\end{references}
\end{document}